\documentclass{wlscirep}
\usepackage[utf8]{inputenc}
\usepackage[T1]{fontenc}
\usepackage{lineno}
\usepackage{makecell}
\usepackage{multirow}
\usepackage{amsmath}
\usepackage{graphicx} 
\usepackage{subfig}   
\usepackage{caption} 
\usepackage{tablefootnote}

\title{A dataset for cyber threat intelligence modeling of connected autonomous vehicles}

\author[1]{Yinghui Wang}
\author[1,2,3,*]{Yilong Ren}
\author[4,5]{Hongmao Qin}
\author[1]{Zhiyong Cui}
\author[1]{Yanan Zhao}
\author[1,2,3,*]{Haiyang Yu}
\affil[1]{Beihang University, School of Transportation Science and Engineering, Beijing, 102206, China}
\affil[2]{Beihang University, Beijing Key Laboratory for Cooperative Vehicle Infrastructure Systems and Safety Control, Beijing, 102206, China}
\affil[3]{Zhongguancun Laboratory, Beijing, 100094, China}
\affil[4]{Hunan University, College of Mechanical and Vehicle Engineering, Changsha, 410082, China}
\affil[5]{Wuxi Intelligent Control Research Institute of Hunan University, Wuxi, 214115, China}
\affil[*]{corresponding author(s): Yilong Ren (renyilong@buaa.edu.cn) and Haiyang Yu (hyyu@buaa.edu.cn)}

\begin{abstract}
Cyber attacks have become a vital threat to connected autonomous vehicles in intelligent transportation systems. Cyber threat intelligence, as the collection of cyber threat information, provides an ideal approach for responding to emerging vehicle cyber threats and enabling proactive security defense. Obtaining valuable information from enormous cybersecurity data using knowledge extraction technologies to achieve cyber threat intelligence modeling is an effective means to ensure automotive cybersecurity. Unfortunately, there is no existing cybersecurity dataset available for cyber threat intelligence modeling research in the automotive field. This paper reports the creation of a cyber threat intelligence corpus focusing on vehicle cybersecurity knowledge mining. This dataset, annotated using a joint labeling strategy, comprises 908 real automotive cybersecurity reports, containing 3678 sentences, 8195 security entities and 4852 semantic relations. We further conduct a comprehensive analysis of cyber threat intelligence mining algorithms based on this corpus. The proposed dataset will serve as a valuable resource for evaluating the performance of existing algorithms and advancing research in cyber threat intelligence modeling within the automotive field.
\end{abstract}
\begin{document}

\flushbottom
\maketitle

\thispagestyle{empty}


\section*{Background \& Summary}
The emergence of connected autonomous vehicles (CAVs) is considered a significant technological breakthrough in the global transportation industry. It is expected that CAVs will greatly improve transportation safety by enhancing efficiency, reducing congestion and minimizing accidents, etc \cite{gupta2022novel,bendiab2023autonomous,mansourian2023deep}. Despite the progress made in advanced automation, enhanced connectivity, and the implementation of shared services, these trends have also brought about new cyber-risks and vulnerabilities. Hackers now have more opportunities to exploit these potential attack surfaces and even gain vehicle control \cite{maple2019connected,pandey2022review}. What's worse, various emerging features such as the remote control function, over-the-air (OTA) update, internet connectivity, external devices and charging infrastructure have transformed potential threats into an undeniable reality. Over the past decade, the frequency, magnitude and sophistication of cyber-attacks on CAVs have experienced an exponential increase \cite{upstream}. These cyber-attacks may cause privacy disclosure, financial losses, human damage or even compromise national public safety \cite{wang2022automotive}. The homologation of new vehicle types now mandates the monitoring and response to cyber-attacks on vehicles and their ecosystem, as per the recent UNECE legal regulation \cite{burkacky2020cybersecurity}. There are numerous publications that discuss various security measures in CAVs, including access control, firewall, intrusion detection and prevention system (IDPS), security operations center (SOC), as well as cyber resilient architectures, etc \cite{wu2019survey,luo2019cyberattacks,ring2018cybersecurity,CyReV,grimm2021context}. Nonetheless, these methods have several limitations, such as passive protection, restricted capability of threat identification, and so forth. Fortunately, cyber threat intelligence (CTI) becomes an ideal way to realize proactive defense and timely response to unknown or emerging threats in the automotive field. However, the mining and analysis of CTI data requires a great deal of manual inspection from open-source unstructured texts, which is an exceedingly time-consuming task \cite{zhao2020cyber}. Automatically extracting CTI knowledge from massive amounts of unstructured data has become a pressing and critical topic in the automotive cybersecurity domain. 
\par Research on CTI knowledge extraction or modeling is still in the early stages, typically relying on pipeline methods, which first perform named entity recognition and then extract relationships between entities \cite{wu2020effective,qin2019network,wang2021method}. Nevertheless, these pipeline methods may result in error propagation in entity recognition, and disregard the correlation between entity recognition and relation extraction. Increasingly, the end-to-end joint extraction model is being recognized as a popular approach for CTI mining tasks. Li et al. developed an end-to-end entity-relation extraction method, incorporating Luo's joint labeling scheme \cite{luo2020neural,li2020knowledge}. The researchers merged the BiLSTM-LSTM model with a dynamic attention mechanism, i.e., BiLSTM-dynamic-att-LSTM model, achieving the simultaneous extraction of both entities and their relations. Subsequently, Zuo et al. constructed an extraction framework based on BERT-BiLSTM-att-CRF \cite{zuo2022end}. Similarly, this work has introduced a sequence label approach and combined extraction principles, enhancing the capability to recognize overlapping relationships. Moreover, the BERT model takes advantage of deep bidirectional transformers to form word embeddings by fusing contextual details. This capability empowers it to effectively capture intricate semantic features during the extraction process \cite{devlin2019bert}. Guo et al. presented the CyberRel, a joint model for extracting entities and relations in cybersecurity concepts using the Bert-BiGRU-att-BiGRU-CRF scheme \cite{guo2021cyberrel}. They modeled the information extraction problem as a multi-sequence tagging process, resulting in distinct label sequences for various relations. Nevertheless, the complexity of this joint model correspondingly increased due to the multiple-sequence labeling strategy. Data is an essential foundation for research in CTI modeling. Mittal et al. proposed a ``CyberTweets" dataset, which was built through tweets collected from Twitter \cite{mittal2016cybertwitter}. The CyberTweets only contained 7 entity categories: vulnerability, ransomware, DDoS, data leak, general, 0 Day and botnet. Similarly, the ``Cyberthreat" corpus presented by Dionísio et al. was also based on Twitter's data, which also only included 5 entity types due to limited training data \cite{dionisio2019cyberthreat}. Zhao et al. collected CTI data from security blogs, hacker forums, etc., and labeled these texts by ``B-I-O" method \cite{zhao2020cyber}. The dataset primarily focused on the regular indicator of compromise, overlooking other complex related entities and relationship types. In addition, Satyapanich et al. annotated a cybersecurity event dataset, which covered cyber attacks and vulnerability information \cite{satyapanich2020casie}. They defined 5 event types, including 14 semantic roles and 20 argument types. Despite the existence of several open-source datasets for CTI modeling, none are specifically focused on CTI entity and relation mining in the automotive domain. What's more, the CAVs are filled with numerous cybersecurity entities. These range from the electronic control unit (ECU), data, in-vehicle network, to its various functions, as well as the potential threats and vulnerabilities it faces. The composition of these cybersecurity entities is irregular, and they face challenges to semantic heterogeneity. In addition, there are many overlapping relational entities in the automotive CTI texts, i.e., one entity may be involved in several semantic relationships simultaneously. These overlapping relations might cause ambiguities, thus making it difficult to recognize such entities and relationships during CTI data mining. Our objective is to provide data support for CTI modeling by developing an automotive CTI ontology and simultaneously annotating security entities and their relationships. This research may provide a solution for effectively extracting security-safety knowledge hidden within vast amounts of cyber threat information, thereby enabling timely detection of potential cyber threats for CAVs.
\par To address the lack of data on automotive CTI knowledge extraction and promote research in CTI modeling, we built an automotive CTI dataset, Acti, focusing on mining CTI entities and their associations. This dataset can be used for CTI modeling, enabling the timely identification and analysis of potential cyber threats to vehicles. This dataset contains data from 908 real cybersecurity incident texts collected across three cyber threat information sources. The dataset includes 10 entity concepts related to cyber and physical world, along with 10 semantic relationship categories. These entities and relationships are derived from the definition of the automotive CTI ontology. The CTI ontology is a way used to formally express the concepts and their semantic relationships in the automotive cyber threat intelligence domain. We have adjusted our data to be the "BIOES" - "entity type" - "relation type" - "entity role" joint annotation schema, and maded the data available at Github under the \hyperlink{https://github.com/AutoCS-wyh/Automotive-cyber-threat-intelligence-corpus}{https://github.com/AutoCS-wyh/Automotive-cyber-threat-intelligence-corpus}. This dataset includes 3678 sentences, covering 8195 security entity instances and 4852 entity-relation triples. We train two CTI mining models using entity-relation joint extraction techniques to validate the reliability of the CTI dataset. Besides, the data was categorized into cyber and physical elements in accordance with the CTI ontology, further depicting and analyzing the interrelation between security and safety. This CTI dataset is expected to facilitate the collaborative analysis of functional safety and cybersecurity, enabling supporting further vehicle cybersecurity research work. In this study, we aim to widely introduce this CTI dataset and make it accessible for public use, enabling more people to conduct meaningful investigations. 


\section*{Methods}

In this section, we present an overview of the data collection and processing steps involved in producing the Acti corpus. The methodology comprises two main components: (1) data collection, which involves identifying and recording cybersecurity attack incidents that are directly related to automobiles; and (2) data processing, which includes ontology modeling and converting the raw cybersecurity incidents data from unstructured descriptive information into "BIOES" - "entity type" - "relation type" - "entity role" joint annotation data.



\subsection*{Data collection}
We mainly collect vehicle cybersecurity data through the following two channels: (1) retrieving published vehicle-related cybersecurity vulnerability information from the national vulnerability database (NVD); and (2) collecting cybersecurity data from specific vehicle threat intelligence platforms, cybersecurity conferences, reports and literature, etc. Currently, the NVD online database has collected some cybersecurity vulnerability data related to CAVs. This data is stored along with over 200,000 vulnerability entries from other industries. Hence, we use keywords retrieval to search for vehicle vulnerabilities in the NVD. According to our previous work \cite{wang2022automotive}, these keywords can be divided into three types, as shown in Table \ref{tab:table1}. 

\begin{table}[ht]
\centering
\begin{tabular}{|c|l|}
\hline
\textbf{Keyword type} & \multicolumn{1}{c|}{\textbf{Keyword}} \\
\hline
Vehicle term & Vehicle, Car, Automotive\\
\hline
Vehicle components and networks & Bluetooth, Braking system, Engine control, Infotainment, Keyless entry,\\~&CAN, LIN, MOST, OBD-II, T-BOX, Gateway, TPMS, etc. \\
\hline
Original equipment manufacturer (OEM) & Bmw, Audi, Toyota, Jeep, Mercedes-benz, Ford, Mazda, Subaru, Great \\~&wall, Lexus, Chrysler, Tesla, etc.\\
\hline
\end{tabular}
\caption{\label{tab:table1} Vehicle cybersecurity data retrieval keywords}
\end{table}

\par In addition, we also collect vehicle cybersecurity data from the auto-threat intelligence cyber incident repository of Upstream Security and automotive attack database (AAD) as provided through Sommer et al. \cite{sommer2019survey}. The website links are \hyperlink{https://upstream.auto/research/automotive-cybersecurity} {https://upstream.auto/research/automotive-cybersecurity}
and \hyperlink{https://github.com/IEEM-HsKA/AAD}{https://github.com/IEEM-HsKA/AAD}. Upstream Security's repository displays more than 1,300 publicly reported cyber incidents specifically targeted at the smart mobility ecosystem. Our focus is exclusively on analyzing the threat or vulnerability information that directly or indirectly impacts vehicles. Therefore, we exclude any ransomware attacks, information leakages or other security incidents that occurred in manufacturers backend platforms, car rental platforms, and more. Meanwhile, we only gather the description information concerning vehicle vulnerabilities and auto-threat intelligence cyber incidents, without taking into account any other field contents. Besides, the AAD contains 23 fields mapping with the attack taxonomy like the year, description, attack class, attack type and vulnerability, etc. Similarly, we select the three columns of unstructured text, i.e., description, attack path and consequence information. On the whole, we collect a total of 908 real automotive cybersecurity data. A brief overview of the essential data contained in the Acti corpus is presented in Table \ref{tab:table2}. 

\begin{table}[ht]
\centering
\begin{tabular}{|c|c|c|}
\hline
\textbf{Data source} & \textbf{Count of CTI} & \textbf{Count of sentences}\\
\hline
Automotive CVE vulnerability (NVD) & 198 & 380\\
\hline
Upstream's cybersecurity incident & 360 & 1219 \\
\hline
Automotive attack database & 350 &2079\\
\hline
Total & 908 &3678\\
\hline
\end{tabular}
\caption{\label{tab:table2} The Overview of Acti corpus}
\end{table}


\subsection*{Data processing}
The goal of the data processing is to label the entity categories and relation types from the unstructured cybersecurity data, and convert them into the "BIOES" - "entity type" - "relation type" - "entity role" joint annotation format. The first step in the data processing workflow is defining a vehicle CTI ontology to describe the entity categories and their interrelations. Next, the data annotation process is performed manually using the brat tool. Brat is a web-based tool for text annotation, used for adding notes to existing text documents. This open-source tool may be downloaded at the following this link \hyperlink{https://github.com/nlplab/brat/archive/refs/tags/v1.3p1.tar.gz}{https://github.com/nlplab/brat/archive/refs/tags/v1.3p1.tar.gz}. Once the manual annotation process is finished, the brat tool would automatically generate “.ann” and “.conll” files. The “.conll” file is only available in the standard BIO format. Thus, we develop a python script to convert all “.ann” files into the joint annotation format. 

\subsubsection*{Ontology modeling for automotive CTI}
\begin{figure}[ht]
\centering
\includegraphics[width=6in]{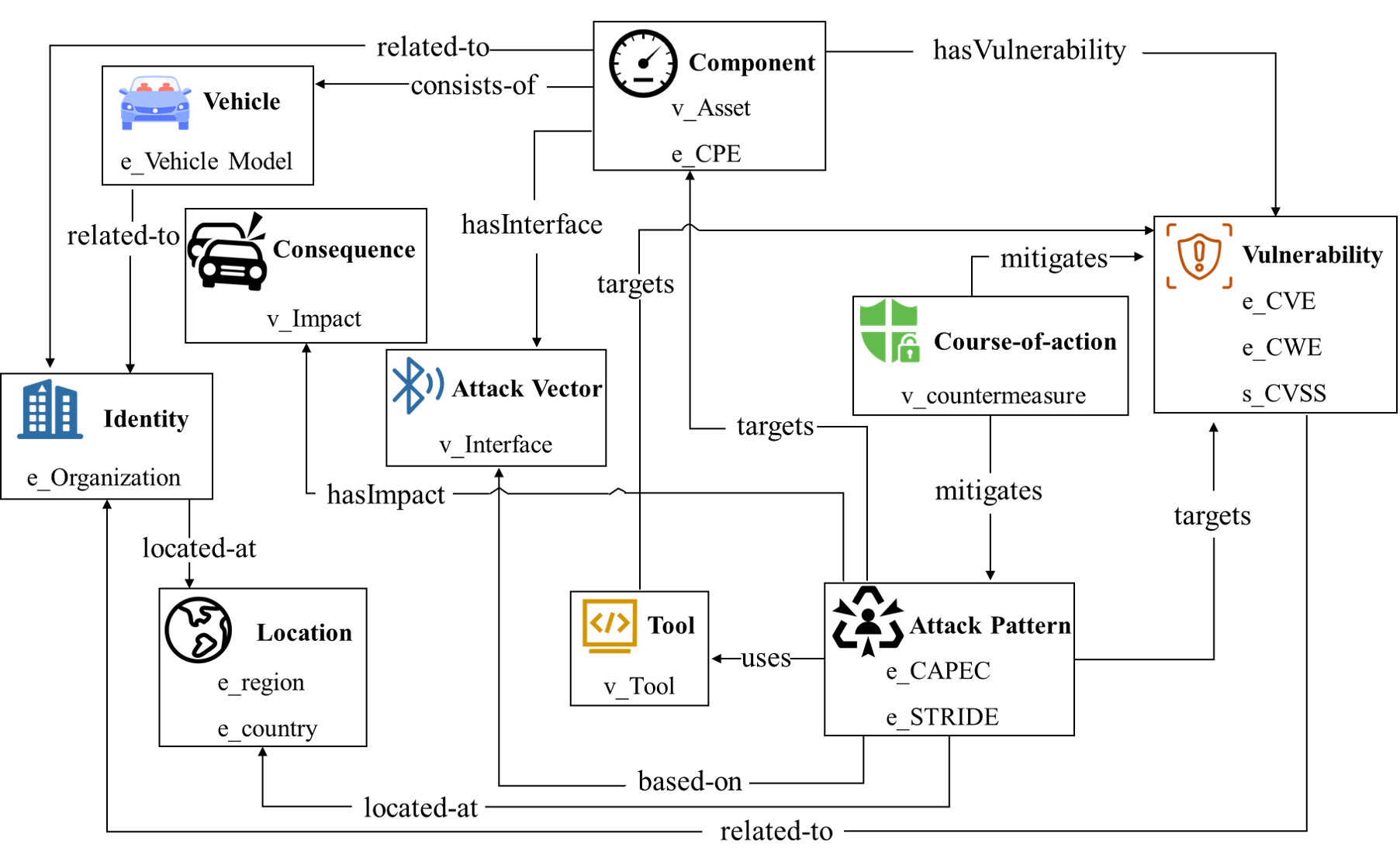}
\caption{Automotive CTI Ontology Model.}
\label{fig1}
\end{figure}

Most cyber threat intelligence is unstructured or encoded in diverse formats, with dispersed data sources, numerous concepts and complex semantic relationships. The ontology concept provides a feasible solution to the problems of heterogeneous data and complex characteristics in the threat intelligence field \cite{menges2019unifying}. During the process of corpus construction, it is crucial to establish a well-defined ontology model for automotive CTI. To provide a comprehensive depiction of CTI in the automotive domain, we define 10 entity and 10 relationship types. These contents integrate industry-leading outcomes, such as unified cybersecurity ontology (UCO) \cite{syed2016uco}, structured threat information eXpression (STIX), the classification model for automotive cyber-attacks \cite{sommer2019survey} and the CTI data model \cite{menges2019unifying}. The automotive CTI ontology model is shown in Fig. \ref{fig1}. It defines 10 entities as follows:

\begin{itemize}
\item[$\bullet$] Component: The component (or asset) class refers to any physical or logical element of a vehicle system or network, including hardware, software, firmware, data and interfaces, etc.
\item[$\bullet$] Consequence: This class describes the potential impact of an attack, such as vehicle theft, control vehicle systems, data breach, service/business disruption, location tracking, fraud, and so on.
\item[$\bullet$] Identity: The identity class represents individuals, organizations or groups, e.g., Acura, BMW, Chrysler, Ford, and so forth.
\item[$\bullet$] Vehicle: The class refers to the vehicle model, which is a distinctive identifier given by the vehicle manufacturer to a particular type of vehicle. e.g., the Tesla Model S. 
\item[$\bullet$] Location: The region where the attack occurred or the target organization/company belongs to.
\item[$\bullet$] Attack vector: The different points where an attacker might enter or retrieve data in a system, often referred to as the attack surface. This class consists of sub-classes like cellular, Bluetooth, CAN-bus, and so on.
\item[$\bullet$] Attack pattern: This class describes various ways attackers employ to compromise targets and consists of sub-classes like brute force, replay, eavesdropping, buffer overflow, reverse engineering, etc.
\item[$\bullet$] Tool: This class represents the equipment that could be employed by attackers to execute an attack. The tools include vehicle diagnostic tools, debuggers, sniffing tools, and so forth.
\item[$\bullet$] Vulnerability: A Vulnerability is an internal mistake that enables an external threat to compromise the system causing security consequences. For example, missing input validation fault, buffer overflow vulnerability, etc.
\item[$\bullet$] Course of action: The course of action is an action that is taken to prevent or respond to an attack, such as access control, encryption, patch, firewall, and so forth.
\end{itemize}

Afterwards, considering the specific attributes of automotive CTI data and the discoveries from previous research, the ontology model incorporates 10 relation types to depict the connections among the predefined threat entities. Specifically, the relation types are: "hasVulnerability", "hasInterface", "hasImpact", "targets", "uses", "mitigates", "related-to",  "located-at", "based-on" and "consists-of". The semantic relations between entity classes are reflected through object properties, as depicted in Table \ref{tab:table3}. These would help to enhance the understanding of how aforementioned pre-defined entities are interconnected. Furthermore, given the intricate and professional nature of the automotive cybersecurity sector, we developed a comprehensive dictionary encompassing specialized vocabulary and public enumeration. This dictionary resource serves as a valuable tool for precisely annotating automotive CTI data. Below, we offer a brief overview of the dictionary.

\begin{table}[htbp]
\caption{ACTI entity-relation triples\label{tab:table3}}
\centering
\begin{tabular}{|c|l|c|}
\hline
\textbf{Relation type} & \multicolumn{1}{c|}{\textbf{Property}}& \textbf{Triple}\\
\hline
hasVulnerability & a specific component has a vulnerability instance& $(Component, hasVulnerability, Vulnerability)$\\
\hline
hasInterface &a component instance has an attack vector instance& $(Component, hasInterface, Attack\ vector)$\\
\hline
hasImpact &\makecell[l]{the attack pattern or vulnerability can cause some \\impact directly or indirectly}& $\makecell{(Attack \ pattern, hasImpact, Consequence) \\( Vulnerability, hasImpact, Consequence)}$\\
\hline
targets &\makecell[l]{the attack pattern/tool targets the component, or\\ vulnerability}& $\makecell{(Attack \ pattern, targets, Component) \\(Attack\ pattern, targets, Vulnerability) \\(Tool, targets, Vulnerability)}$\\
\hline
uses &\makecell[l]{the tool is employed to carry out the behavior\\identified in the attack pattern}& $(Attack \ pattern, uses, Tool)$\\
\hline
mitigates &\makecell[l]{the course of action can mitigate the related attack \\pattern or vulnerability}& $\makecell{(Course \ of\ action, mitigates, Attack \ pattern) \\ (Course\ of\ action, mitigates, Vulnerability)}$\\
\hline
related-to &two entity classes have a related relationship& $\makecell{(Component, related\text{-}to, Identity) \\(Vehicle, related\text{-}to, Identity) \\(Vulnerability, related\text{-}to, Identity)}$\\
\hline
located-at & \makecell[l]{the identity is located at the related location, or an\\ attack occurred in a region}&$\makecell{(Identity, located\text{-}at, Location) \\(Attack \ pattern, located\text{-}at, Location)}$\\
\hline
based-on & \makecell[l]{the attack pattern instance was carried out through \\the attack vector}&$(Attack \ pattern, based\text{-}on, Attack\ vector)$\\
\hline
consists-of & two entity classes have a containment relationship& $\makecell{(Component, consists\text{-}of, Vehicle) \\(Component, consists\text{-}of,Component)}$\\
\hline
\end{tabular}
\end{table}

\begin{itemize}
\item Vocabulary: The vocabulary represents the list of candidate content for pre-defined entities that is supplied with automotive CTI ontology, namely internal enumeration. The vocabulary list mainly includes "asset", "impact", "interface", "countermeasure" and "tool" categories. For example, the asset vocabulary contains more than 1,000 terms, such as central gateway, brake system ECU, advanced driver assistance system (ADAS), etc. The tool vocabulary list mainly consists of hardware, software, security, sensing, measurement, and wireless tool, with previous reference to the research of Sommer et al. \cite{sommer2019survey}.
\item Public Enumeration: The public enumeration refers to publicly available expressions or databases that contain valuable data related to CTI. This includes critical details such as the configuration, weakness, vulnerability, vehicle type, location and among other aspects. The exemplary enumerations involve CVE number, common weakness enumeration (CWE), common attack pattern enumeration and classification (CAPEC), and so forth.
\end{itemize}

\subsubsection*{Automotive CTI modeling corpus annotation}

\begin{figure}[ht]
\centering
\includegraphics[width=4.6in]{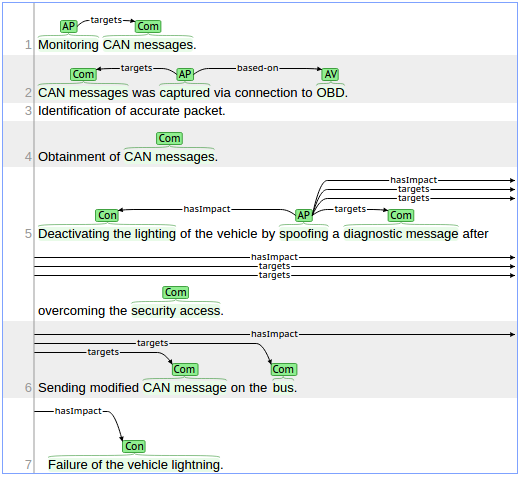}
\caption{Brat manual annotation.}
\label{fig2}
\end{figure}

We integrate the entity and relation labels into the joint annotation scheme \cite{luo2020neural}, which enables the information among them to be fully exploited. This tagging scheme contains four parts, i.e., entity boundary, entity type, relation type and entity role. There are three label types of tokens: (1) tokens unrelated to either entities or the relations, (2) tokens strictly related to entities, and (3) tokens that encompass both entities and relations. Therefore, the number of labels is $N = 3*4*e*(r+1)+ 4*e+1$, where e, r are the size of the entity and relation type set, respectively. The detailed scheme is as follows:
\begin{itemize}
\item Entity boundary: The "BIOES"  (Begin, Inside, Other, End, Single) scheme is adopted to indicate the token's position within the entity. The "B", "I", and "E" tags denote the position of the word, i.e., the beginning, middle and tail of the entity, respectively. The tag "S" represents that the word is a single entity. Additionally, the tag "O" means that the word isn't associated with any pre-defined entities.
\item Entity type: The entity type information is pre-defined according to the automotive CTI ontology model. We classify entities into ten categories: component (Com), consequence (Con), identity (Ide), vehicle (Veh), location (Loc), attack vector (AV), attack pattern (AP), tool (Tool), vulnerability (Vul) and course of action (CoA).
\item Relation type: The relation type is also acquired from the pre-defined set: \{"hasVulnerability", "hasInterface", "hasImpact", "targets", "uses", "mitigates", "related-to", "located-at", "based-on" , "consists-of"\}. Furthermore, the tag "M" is added to denote the entity exits multiple relations, i.e., overlapping relations.
\item Entity role: The label indicates the role of the entity in the relation and is defined using the tags "1", "2" and "m". Concretely, the labels "1" and "2" represent the word of the first entity and second entity in the relation, respectively. The "m" represents the word of distinct role entities in the overlapping relation.
\end{itemize}
After determining the entity and relation types that need to be annotated, the data annotation process begins. The process is known as a sequence labeling task. Annotating each element in the open-source unstructured data with a label by considering each sentence as a sequence, where each word is an element. We leverage the above annotation strategy and the open-source brat tool to effectively label entity and relation types in automotive cybersecurity data. Before data annotation, the brat annotation tool requires the definition of the annotation format first (manually defining entity and relationship types through the brat tool configuration file). Then, the brat annotation tool is used to label various types of entities. And the connection between two entities is established to annotate the relation type. The entity and relation annotation is completed manually, as shown in Fig. \ref{fig2}. 

\begin{figure}[ht]
\centering
\includegraphics[width=4.6in]{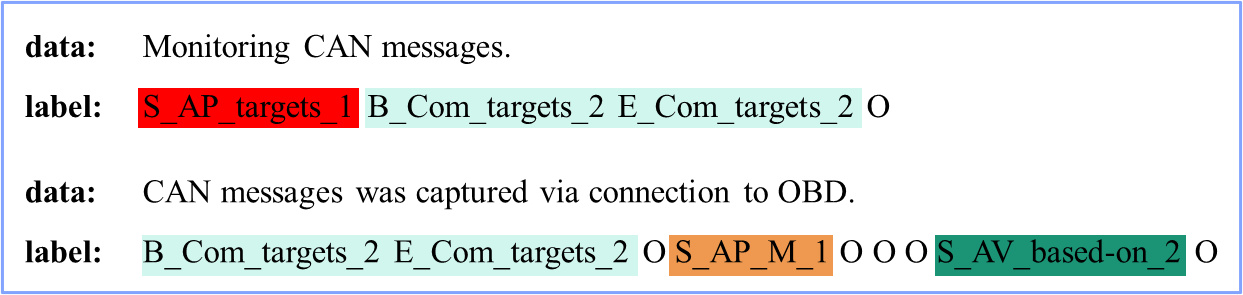}
\caption{Automotive CTI annotation data.}
\label{fig3}
\end{figure}

Once the manual annotation process is finished, the brat tool could automatically generate ".ann" and ".conll" files. However, the ".conll" file is only available in the standard BIO format. Thus, we have developed a conversion script to translate these ".ann" files into the joint annotation format described above. Take the sequence of "Monitoring CAN message.", the BIO format is "S-AP B-Com E-Com O", and the corresponding joint annotation label would be represented as "S-AP-targets-1 B-Com-targets-2 E-Com-targets-2 O", as shown in Fig. \ref{fig3}. 

\section*{Data Records}
The Acti dataset is available on GitHub (\hyperlink{https://github.com/AutoCS-wyh/Automotive-cyber-threat-intelligence-corpus}{https://github.com/AutoCS-wyh/Automotive-cyber-threat-intelligence-corpus}). This dataset contains raw unstructured cybersecurity data (.txt files), annotation data files using brat tool (.ann files) and joint annotation data (.txt files). The raw unstructured cybersecurity data is collected from NVD, auto-threat intelligence cyber incident repository of Upstream Security and automotive attack database on GitHub. The raw dataset includes 908 files in ".txt" format, with names such as "CVE-2022-23126", "2022-11-1" and "ID2019NxumaloSSA1", corresponding to the aforementioned data sources. To facilitate the usability of the dataset and ensure reproducibility, we present detailed instructions in the README file, describing the Acti dataset processing workflow. Besides, we randomly divided the entire Acti dataset into training and test sets in a ratio of 8:2. The source code to convert these ".ann" files into the joint annotation format and preprocess the joint annotation data can be found in this GitHub repository. Furthermore, the source code for training CTI knowledge mining deep learning models is also available and will be provided in the Code Availability section.



\section*{Technical Validation}
To evaluate the performance of this Acti corpus, we compared various open source corpora, and conducted a comprehensive analysis of deep learning algorithms for CTI knowledge extraction. Specifically, we evaluated the performance of the end-to-end CTI joint extraction models on the Acti dataset. These methods avoid the error propagation inherent in the classical pipeline model and have increasingly become a key direction in CIT modeling research.

\subsection*{Comparison of corpora}
We carefully investigated various open-source corpora for CTI knowledge mining tasks in the cybersecurity domain, as shown in Table \ref{tab:table4}.
The "Cybertweets", "Cyberthreat" and "HINTI" corpora only provide the annotation for entity types, regardless of the semantic relations among security entities. Furthermore, the entity types labeled in these datasets are so restricted that they inadequately capture the comprehensive landscape of CTI information. The "CASIE" dataset contains annotations for both entities and relations. However, it simply annotated the entity type while ignoring the labeling of entity boundaries. And the problem of overlapping relational entities was also not considered. What's more, none of the above datasets cover information related to automobile cybersecurity. In this paper, we employ the sequence tagging strategy to annotate entities and relations within the automotive CTI data, effectively alleviating the problem of overlapping relational entities. The Acti corpus is a valuable source of automotive CTI information mining tasks, offering a comprehensive view of CTI entities and relations. Besides fundamental security-related elements, the Acti also incorporates the marking of vehicle components, physical impact and other entities. These elements significantly facilitate the study of the interrelation between automobile cybersecurity and functional safety. 

\begin{table}[htbp]
\caption{Cybersecurity knowledge extraction open source corpora\label{tab:table4}}
\centering
\begin{tabular}{|p{3.5cm}<{\centering}|p{1.8cm}<{\centering}|p{1.8cm}<{\centering}|p{1.8cm}<{\centering}|}
\hline
\textbf{Corpus} & \textbf{Entity} & \textbf{Relation}& \textbf{Scale}\\
\hline
CASIE\tablefootnote{https://github.com/Ebiquity/CASIE.} & 20 & 14 & 10384\\
\hline
Cybertweets\tablefootnote{https://github.com/behzadanksu/cybertweets.} & 8 & - & 21000\\
\hline
Cyberthreat\tablefootnote{https://github.com/ndionysus/twitter-cyberthreat-detection.} & 5 & - & 11073 \\
\hline
HINTI\tablefootnote{https://cse.msu.edu/$ \sim $qyan/CTIdataset\_Release.zip.} & 6 & - & 30000 \\
\hline
\textbf{Acti} & \textbf{10} & \textbf{10} & \textbf{3678} \\
\hline
\end{tabular}
\end{table}

\subsection*{CTI knowledge mining models}
\begin{figure}[htbp]
\centering
\includegraphics[width=4.16in]{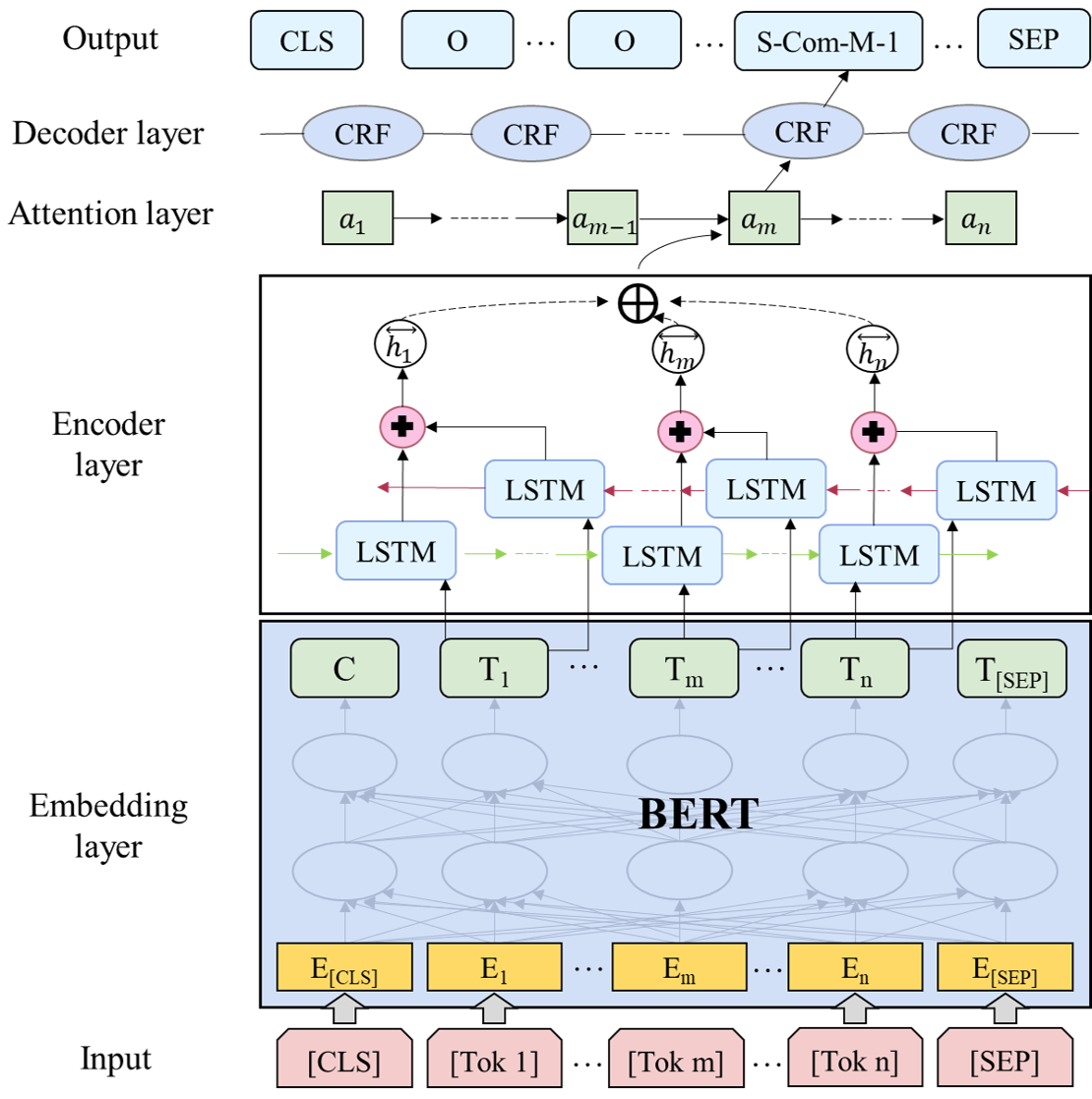}
\caption{BERT-BiLSTM-att-CRF model.} 
\label{fig4}
\end{figure}

Considering the annotation scheme of Acti dataset, the "BERT-BiLSTM-att-CRF" \cite{zuo2022end} and "BiLSTM-dynamic-att-LSTM" \cite{li2020knowledge} models are employed to verify the reliability of the Acti corpus for automotive CTI knowledge mining tasks. The architecture of the BERT-BiLSTM-att-CRF model is illustrated in Fig.\ref{fig4}. This model consists of four layers: embedding, BiLSTM, attention mechanism and conditional random fields (CRF) \cite{zuo2022end}. The embedding layer transforms words into vector representations, and then the BiLSTM layer calculates the probability distribution of the word vector. Subsequently, the attention layer reflects the relationships between words, and the CRF layer predicts the globally optimal labeling sequence.

\begin{itemize}
\item \textbf{Embedding layer}.  The BERT model encodes the inputs on multiple transformer layers, incorporating multi-head attention and feedforward neural networks to effectively extract deep semantic information. 
\item \textbf{Encoder layer}. The BiLSTM, or bidirectional long short-term memory (LSTM), is a highly potent network that utilizes both forward and backward LSTM to effectively capture semantic information in both directions of each sequence. It could address the issue of gradient disappearance and enhance the generation of comprehensive semantic features.
\item \textbf{Attention mechanism layer}. The self-attention mechanism is introduced to focus on the key information from the cybersecurity dataset. It initializes an attention matrix to represent the relative significance of words to their vectorization representations, and effectively reflects the influence between words. 
\item \textbf{Decoder layer}. Labels exhibit strong interconnections rather than being independent. The CRF model is commonly used in sequence labeling tasks. It effectively calculates the transfer probability between labels in a tagging sequence, allowing it to select a global optimal sequence label.
\end{itemize}

\begin{figure}[htbp]
\centering
\includegraphics[width=4.16in]{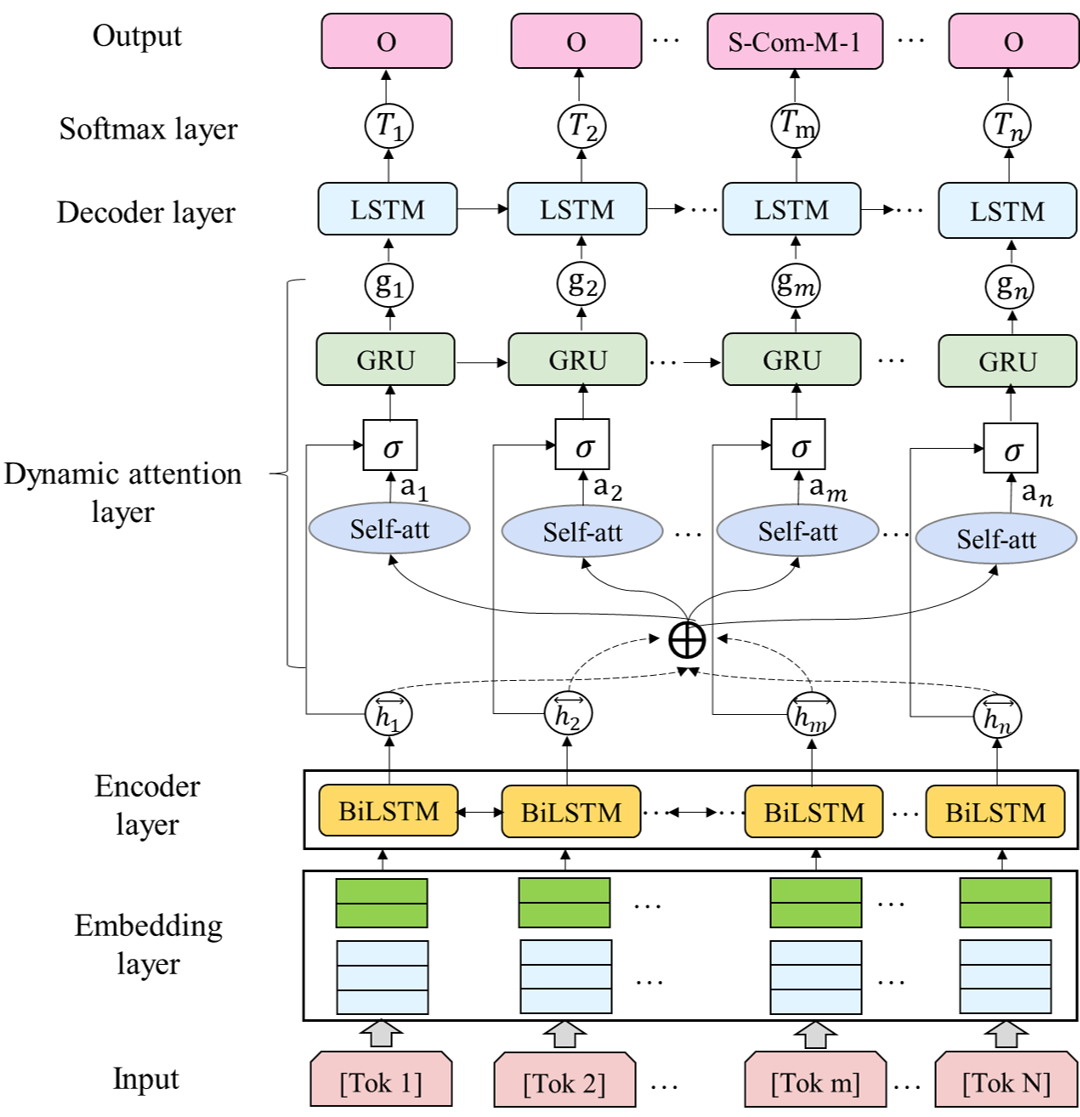}
\caption{BiLSTM-dynamic-att-LSTM model.} 
\label{fig5}
\end{figure}

The structure of the BiLSTM-dynamic-att-LSTM model is shown in Fig.\ref{fig5}. This model includes the embedding layer, encoder layer, dynamic attention layer, decoder layer and softmax layer.

\begin{itemize}
\item \textbf{Embedding layer}. The word2vec is used to obtain word-level features of the sentence sequence. Then, the convolutional neural network (CNN) module is adopted to extract the character features of the input sentence sequence. Finally, these two features are spliced together as the input of the encoding layer.
\item \textbf{Encoder layer}. The encoder layer adopts the same method as the BERT-BiLSTM-att-CRF model, i.e., BiLSTM, to obtain the feature encoding of input sequences.
\item \textbf{Dynamic Attention Mechanism layer}. The dynamic attention mechanism takes into account the semantics of words and considers their variations in different contexts \cite{li2020knowledge}. The feature encoding $h_t$ generated by the BiLSTM layer is spliced with the output $a_t$ of the self-attention mechanism. The splicing result [$h_t$, $a_t$] is filtered by the sigmoid function to obtain $\gamma_t$, then a dot-product operation is performed to calculate $\xi_t$. The $\xi_t$ is the input for the gated recurrent unit (GRU). The detailed process is as follows:
\begin{equation}  
	\gamma_t = sigmoid(W_s [h_t, a_t]) 
\end{equation}
\begin{equation}  
	\xi_t = \gamma_t [h_t, a_t]
\end{equation}
\begin{equation}  
	g_t = GRU(g_{t-1},\xi_t,\theta)
\end{equation} 
where $W_s$ and $\theta$ are hyper-parameter matrices. Let $G=\{g_1,g_2,...,g_t\}$ represent the final output of the dynamic attention layer, and subsequently input to the lower layer for decoding.
\item \textbf{Decoder layer}. The decoding layer employs a LSTM network to generate the vector representation of label sequences. The LSTM tag decoder can significantly speed up model training and promises to achieve comparable performance to the CRF decoder. Finally, the label sequence is normalized using a softmax layer. 
\end{itemize}

\subsubsection*{Pre-processing and experimental setting}
This specialized cybersecurity dataset for automotive cyber threat intelligence modeling comprises 3678 sentences, 8195 labeled entities and 4852 relationships. we randomly divided the entire Acti dataset into training and test sets in a ratio of 8:2. The experiments are carried out on a high-performance server running the Windows 10 operating system. And the server is equipped with an Intel Core i5-13400F CPU@2.50 GHz, 64 GB RAM and the powerful NVIDIA GeForce RTX 3090 GPU. The software environment comprised Python 3.7, CUDA 11.2, PaddlePaddle-GPU 2.3.2 and paddlenlp 2.1.1, etc. Some main hyperparameters of the "BERT-BiLSTM-att-CRF" and "BiLSTM-dynamic-att-LSTM" models are shown in Table \ref{tab:table5}.

\begin{table}[htbp]
\caption{Hyperparameter setting\label{tab:table5}}
\centering
\begin{tabular}{|p{5cm}<{\centering}|p{4cm}<{\centering}|p{2cm}<{\centering}|}
\hline
\textbf{Model} & \textbf{Parameter} & \textbf{Value} \\
\hline
\multirow{10}{*}{BERT-BiLSTM-att-CRF} & Transformer layers  & 12\\
\cline{2-3}
~ & hidden size & 768\\
\cline{2-3}
~ & activation function  & ReLU \\
\cline{2-3}
~  & max position embedding  & 128 \\
\cline{2-3}
~  & BiLSTM\_dim & 800 \\
\cline{2-3}
~  &epoch & 85 \\
\cline{2-3}
~  & batch & 16 \\
\cline{2-3}
~  & learning rate & 5e-5 \\
\cline{2-3}
~  & dropout & 0.1 \\
\hline
\multirow{7}{*}{BiLSTM-dynamic-att-LSTM} & word embedding\_dim  & 300\\
\cline{2-3}
~ & char embedding\_dim & 15\\
\cline{2-3}
~ & CNN filter  & 15 \\
\cline{2-3}
~  & CNN kernel size & 5 \\
\cline{2-3}
~  & BiLSTM\_dim & 300 \\
\cline{2-3}
~  &LSTM\_dim & 600 \\
\cline{2-3}
~  & learning rate & 0.001 \\
\cline{2-3}
~  & bias weight & 10 \\
\hline
\end{tabular}
\end{table}

\subsubsection*{Experimental results}
We performed a comprehensive comparison of the Acti dataset for the CTI knowledge mining task. We employ the general evaluation metrics for information extraction tasks, namely precision (P), recall (R) and F1-score, to evaluate the performance of CTI knowledge mining models.

\begin{equation}
\label{1}
P = \frac{TP}{TP+FP}
\end{equation}

\begin{equation}
\label{2}
R = \frac{TP}{TP+FN}
\end{equation}

\begin{equation}
\label{3}
F1 = \frac{2*P*R}{P+R}
\end{equation}

Where TP represents the number of positive instances correctly recognized; FP is the count of positive samples incorrectly identified; and FN stands for the number of negative examples incorrectly classified. The overall performance of these CTI knowledge mining models is evaluated using the aforementioned evaluation indicators. The precision, recall and F1 score are shown in Table \ref{tab:table6}.
\begin{table}[htbp]
\caption{Comparison of experimental results\label{tab:table6}}
\centering
\begin{tabular}{|p{5.5cm}<{\centering}|p{1.5cm}<{\centering}|p{1.5cm}<{\centering}|p{1.5cm}<{\centering}|}
\hline
\multirow{2}{*}{\textbf{Model}} & \multicolumn{3}{c|}{\textbf{Metric}} \\
\cline{2-4}
~& P\% & R\% & F1\% \\
\hline
BERT-BiLSTM-att-CRF & 50.9 & 55.84 & 53.26  \\
\hline
BiLSTM-dynamic-att-LSTM & 47.69 &40.7  & 43.92 \\
\hline
\end{tabular}
\end{table}

We compare the experiment results of the above CTI knowledge extraction models on the Acti corpus. The aforementioned models achieve joint extraction of security entities and their relations, utilizing the semantic relationship between entity recognition and relation extraction tasks. Observably, the BERT-based model outperforms the BiLSTM-dynamic-att-LSTM method, which also proves the exceptional capability of BERT in the CTI knowledge mining tasks.  The reason behind this is that BERT word vector better captures grammatical and semantic information across various contexts, enhancing the model's ability to generalize. Thus, it is capable of effectively handling the complex semantic characteristics of automotive CTI data. Subsequently, we evaluate the performance metrics for each class of entity and relation extraction in the BERT-att-BiLSTM-CRF model. As for entity recognition, a predicted entity is deemed accurate when its boundary and entity category are accurately labeled. Similarly, when the boundary, entity category and relation type are all correct, the relation extraction result is deemed exact. The experiment statistical results are as shown in Table \ref{tab:table7}.

\begin{figure}[htbp]
\centering
\subfloat[{\fontfamily{ptm}\selectfont Entity}]{\includegraphics[width=3.6in]{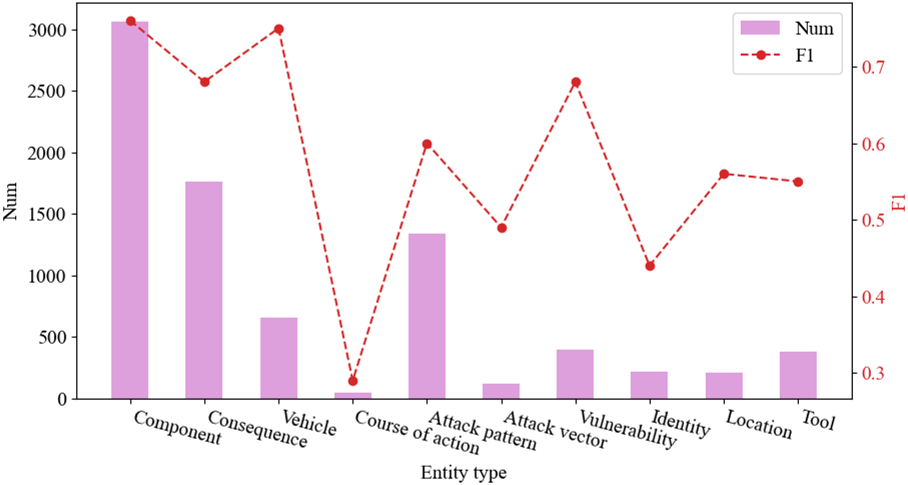}
\label{fig61}}
\hfil
\subfloat[{\fontfamily{ptm}\selectfont Relation}]{\includegraphics[width=3.6in]{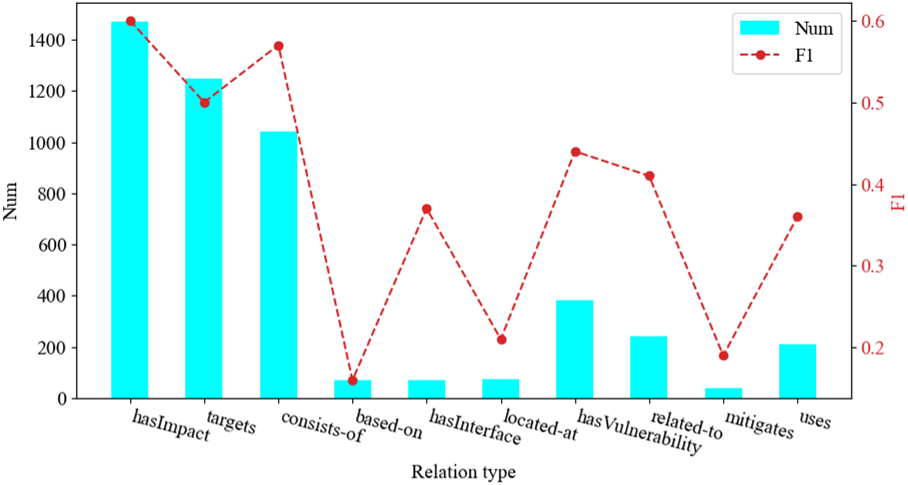}
\label{fig62}}
\caption{Distribution and F1 of entity and relation extraction.}
\label{fig:fig6}
\end{figure}

\begin{table}[htbp]
\caption{Experimental results on entities and relations\label{tab:table7}}
\centering
\begin{tabular}{|p{3cm}<{\centering}|p{1cm}<{\centering}|p{1cm}<{\centering}|p{1cm}<{\centering}|p{3cm}<{\centering}|p{1cm}<{\centering}|p{1cm}<{\centering}|p{1cm}<{\centering}|}
\hline
\textbf{Entity} & \textbf{P\%}  & \textbf{R\%}  & \textbf{F1\%}&\textbf{Relation} &\textbf{P\%}  & \textbf{R\%}  & \textbf{F1\%}\\
\hline
\textbf{Component} & 79.53 & 72.10 & 75.63 & \textbf{hasVulnerability} &62.35 & 33.99 & 44.00 \\
\hline
\textbf{Consequence} & 77.07 & 60.07 & 67.52 &\textbf{hasInterface}& 82.14 & 23.47 & 36.51  \\
\hline
\textbf{Identity} & 60.84 & 34.35 & 43.91 &\textbf{hasImpact} &67.69  &54.67 & 60.48 \\
\hline
\textbf{Vehicle} & 85.65 & 66.78 & 75.05 &\textbf{targets}& 61.39 & 41.80 & 49.73\\
\hline
\textbf{Location} & 66.22 & 48.84 & 56.21 & \textbf{uses} & 51.33 & 27.62 & 35.91\\
\hline
\textbf{Attack vector} & 73.02 & 36.51 & 48.68 &\textbf{mitigates} & 33.33 & 13.33 & 19.05\\
\hline
\textbf{Attack pattern} & 83.13 & 47.03 & 60.07 &\textbf{related-to}& 56.12 & 32.74 & 41.35\\
\hline
\textbf{Tool} &69.43 & 44.97 & 54.59  &\textbf{located-at}&35.56 & 15.24& 21.33\\
\hline
\textbf{Vulnerability} & 81.99 & 58.01 & 67.95&\textbf{based-on} & 26.98 & 11.56 & 16.19\\
\hline
\textbf{Course of action} & 71.88 & 18.25 & 29.11 &\textbf{consists-of} & 65.57 & 50.47 & 57.04 \\
\hline
\end{tabular}
\end{table}
Meanwhile, we further conducted a detailed analysis of the data characteristics within the Acti corpus. The distribution of entity and relation instances is shown in Fig.\ref{fig:fig6}. It can see that there is a clear imbalance in the entity and relation instances of the corpus. This is because the description of original text data mainly focuses on security elements such as component, consequence and attack pattern, etc. The instances of these corresponding relationships also constitute a significant portion in the automotive CTI data. Based on this, we analyze the relationship between the experimental results of each entity and relation extraction and their corresponding instance distribution. From Fig. \ref{fig:fig6}, it can be seen that the F1 values of each entity and relation extraction experiment are generally consistent with their instance distributions. The frequency of several entities and relations in automotive CTI corpus is too low, such as "Course of action", "located-at" and "mitigates", etc., resulting in low performance indicators.

Furthermore, there are many cross-sentence relation entities in the Acti corpus, as shown in Fig.\ref{fig2}. Unfortunately, these exsiting CTI knowledge mining models only consider the entity-relation extraction at the sentence level, ignoring the problem of cross-sentence entities. These may have a certain influence on the performance of CTI knowledge extraction experiments. Despite this, these experiments have also sufficiently demonstrated the reliability of the Acti corpus in extracting entities and their relations of automobile CTI data. Researchers can select or improve these approaches according to the specific requirements and restrictions of their projects.


\section*{Usage Notes}
Our proposed Acti corpus supplies a comprehensive collection of real cybersecurity data related to vehicles, encompassing a wide range of security entities, safety entities and semantic relationships. That is to say, the Acti dataset has the potential to serve as a valuable resource for CTI modeling and for enhancing security analysis in the automotive domain. The Acti dataset can be employed to train deep learning models which can automatically extract CTI knowledge—specifically, security and safety entities and their relationships—from massive amounts of unstructured data, thereby enhancing the ability to timely identify potential threats, and aiding in the formulation of appropriate security measures for automobiles. By supplying researchers with access to an automotive CTI entity-relation joint annotation corpus, the Acti corpus facilitates the development and evaluation of more effective CTI modeling or knowledge mining algorithms. The Acti dataset contains both security-related and physical elements. It is expected to facilitate the collaborative analysis of functional safety and cybersecurity, enabling supporting further cybersecurity research work for CAVs.


\section*{Code availability}

All source code for format conversion and preprocessing of the Acti dataset, as well as deep learning training, and the complete Acti corpus have been uploaded to GitHub at \hyperlink{https://github.com/AutoCS-wyh/Automotive-cyber-threat-intelligence-corpus}{https://github.com/AutoCS-wyh/Automotive-cyber-threat-intelligence-corpus}. The GitHub repository also includes the environment file that defines the versions of python, PaddlePaddle-GPU framework, paddlenlp, etc., and a configuration file that specifies the parameters used for data annotation. The code is openly accessible and can be used freely with appropriate attribution.


\bibliography{sample}


\section*{Acknowledgements} 
This work was funded by the National Key Research and Development Project of China (2022YFB4300404).

\section*{Author contributions statement}
Y.W., Y.R., H.Q., Z.C., Y.Z. and H.Y. prepared the manuscript. Y.W. established the data collection and processing methodology, implemented the data processing, and performed data validation experiments. Y.R. and H.Y. provided financial support. All authors reviewed the manuscript.


\section*{Competing interests} 
The authors declare no competing interests.

\end{document}